\documentclass[a4paper,10pt,twoside]{cpc-hepnp}

\usepackage{multicol}
\usepackage{graphicx}
\usepackage{booktabs}
\usepackage{amssymb,bm,mathrsfs,bbm,amscd}
\usepackage[tbtags]{amsmath}
\usepackage{lastpage}

\begin{document}
%\begin{CJK*}{GBK}{song}

\title{Deuteron potential using the noncritical holographic model}

\author{ %
      Song Li-Tao$^{1;1)}$\email{songlt12@zzu.edu.cn} %
\quad Liu Yong-Xin$^{1}$ %
\quad He Jie$^{2}$
}
\maketitle

\address{%
$^1$ Physical Engineering college, Zhengzhou University, Zhengzhou 450001, China\\
$^2$ College of science, Henan University of Technology, Zhengzhou 450001, China\\
}

\begin{abstract}
Deuteron has been studied using a noncritical holographic quantumchromodynamics model constructed in the six-dimensional Anti-de-Sitter supergravity background. The nucleus potential energy is calculated. The binding energy has been estimated as the minimum of the potential energy. The roles of various mesons in producing the deuteron potential energy are discussed in details.
\end{abstract}



\begin{multicols}{2}

\section{Introduction}

The nucleon-nucleon (NN) interaction is the basis for all of nuclear physics. Many nuclear potential and binding energies have been known with high accuracy in experiments, but they can not be predicted with sufficient accuracy using various theoretical models. Quantumchromodynamics (QCD) as the fundamental theory that describes the strong interaction is always followed with interest by many physicists. The basic properties of QCD are the asymptotic freedom and the color-confinement. At high energy scale, QCD perturbation theory is suitable and powerful. At low energy scale, because of its non-perturbative behavior, to do quantitative analysis of low energy strong interaction process is very difficult. So the QCD picture for the nucleon's structure and property is scale-dependent. There are some high-quality one-boson-exchange potentials to describe the empirical scattering data, such as AV18, CD-Bonn, Reid93, etc., but they contain a lot of purely phenomenological parameters\cite{lab1, lab2, lab3, lab4, lab5}.

Holographic QCD theory provides a new effective way to deal with QCD strong coupling problems\cite{lab6, lab7, lab8, lab9, lab10, lab11, lab12, lab13, lab14, lab15, lab16, lab17, lab18, lab19, lab20, lab21, lab22, lab23, lab24, lab25}. Many critical holographic models such as the $SS$ model\cite{lab26, lab27} have been constructed. They describe some features of QCD well. Recently, some noncritical holographic models were introduced as well. One of them is based on the compactified six-dimensional Anti-de-Sitter ($AdS_6$) space-time with a constant dilaton\cite{lab28, lab29, lab30}. Its low energy effective theory is a four-dimensional QCD-like effective theory. In the model, the six-dimensional gravity background is the near horizon geometry of the color $D4$ brane. This model has just one free parameter which is the mass scale of the model, $M_{KK}$, and all the other parameters are derived from model calculations\cite{lab8}. Some of the QCD features have been studied using this model such as the meson spectrum\cite{lab30}. In this paper, we first briefly describe the noncritical $AdS_6$ model and its NN interaction potential. Then, based on the NN potential, the Deuteron as an example is discussed in details.

\section{$AdS_6$ model and NN potential}

In the $AdS_6$ model, the near horizon gravity background at low energy is written as\cite{lab30}
\begin{equation}\label{eq1}
  ds^2=\left(\frac{U}{R}\right)^2(-dt^2+dx_idx_i+f(U)d{\tau}^2)+\left(\frac{R}{U}\right)^2 \frac{dU^2}{f(U)}.
\end{equation}
The new coordinate $\omega$ is introduced to transform the metric to a conformally flat metric\cite{lab8} as follows:
\begin{equation}\label{eq2}
  d\omega = \frac{R^2U^{1/2}dU}{\sqrt{U^5-U^5_{KK}}}.
\end{equation}
So, the DBI action for the $D4$ brane can be written as
\begin{equation}\label{eq3}
\begin{split}
  S^{D4}_{YM}= & -\frac{1}{4}\mu _4(2\pi \alpha')^2\int d^4 x d \omega e^{-\phi} \left(\frac{U(\omega}{R}\right) trF_{mn}F^{mn} \\
  = & - \int d^4xd\omega \frac{1}{4e^2(\omega)}trF_{mn}F^{mn}.
\end{split}
\end{equation}
The energy of a point-like instanton localized at $\omega = 0$ can be obtained as\cite{lab8}
\begin{equation}\label{eq4}
  m^{(0)}_B=\frac{\sqrt{3/2}\ 4\pi^2\mu_4(2\pi\alpha')^2R}{5}N_cM_{KK}.
\end{equation}

Because the baryon is smaller than the effective length of the fifth direction $\sim 1/M_{KK}$, it can be assumed as a point-like object in five dimensions. Thus an effective action for the baryon field can be written as\cite{lab21}
\begin{equation}\label{eq5}
  \begin{split}
  \int d^4xd \omega &  [-i\mathcal{\bar{N}}\gamma ^m D_m \mathcal{N} - im_b(\omega) \mathcal {\bar{N}N} \\
  & + g_5(\omega)\frac{\rho ^2_{baryon}}{e^2(\omega)}\mathcal{\bar{N}} \gamma^{mn}F_{mn}\mathcal{N} ] \\
  & - \int d^4xd\omega \frac{1}{4e^2(\omega)}trF_{mn}F^{mn}.
  \end{split}
\end{equation}
 This action must be reduced to the four dimension. So, the $KK$ mode expansion should be performed. The baryon field is expanded as
\begin{equation}\label{eq6}
  \mathcal{N}_{L,R}(x^{\mu},\omega)=N_{L,R}(x^{\mu})f_{L,R}(\omega),
\end{equation}
Also, at $A_z = 0$ gauge, the gauge field is expanded as
\begin{equation}\label{eq7}
  A_{\mu}(x,\omega)=i\alpha _{\mu} (x) \psi _0 (\omega) + i \beta _{\mu}(x) + \sum _{n} B^{(n)}_{\mu} (x) \psi _{(n)} (\omega).
\end{equation}
Finally, the four-dimensional NN Lagrangian\cite{lab8} is obtained as
\begin{equation}\label{eq8}
  \mathcal {L} _{nucleon} = -i \bar N \gamma ^\mu \partial _\mu N - im_B \bar N N + \mathcal L _{vector} + \mathcal L _{axial},
\end{equation}
where
\begin{equation}\label{eq9}
  \begin{split}
    & \mathcal L _{vector} = -i \bar N \gamma ^\mu \beta _\mu N - \sum _ {k \geq 0} g^{(k)}_V \bar N \gamma ^\mu B^{(2k+1)}_{\mu}N , \\
    & \mathcal L _{axial} = - \frac{ig_A}{2} \bar N \gamma ^\mu \gamma ^5 \alpha _\mu N - \sum _ {k \geq 1} g^{(k)}_A \bar N \gamma ^\mu \gamma ^5 B^{(2k)}_{\mu}N .
  \end{split}
\end{equation}

In the $AdS_6$ model, the leading parts of NN potential come from the pseudoscalar meson $\pi$, isovector vector meson $\rho$,  isoscalar vector meson $\omega$, and isovector axial vector meson $a$ exchange interactions. So, the noncritical holographic NN potential can be written as\cite{lab9, lab10}
\begin{equation}\label{eq10}
  V^{holography}_{NN}=V_C(r)+(V^{\sigma}_T(r)\vec{\sigma} _1 \cdot \vec {\sigma} _2 + V^S_T(r)S_{12})\vec {\tau} _1 \cdot \vec {\tau} _2 ,
\end{equation}
where
\begin{equation}\label{eq11}
  V_C(r)= \sum _{k=1}^{P} \frac{1}{4\pi} (g_{\omega^{(k)}\mathcal{NN}})^2 m_{\omega^{(k)}} y_0(m_{\omega^{(k)}} r),
\end{equation}
\begin{equation}\label{eq12}
\begin{split}
   V_T^{\sigma}(r) = & \sum _{k=1}^{P} \frac{1}{4\pi} \left ( \frac{g_{\rho^{(k)}\mathcal{NN}}M_{KK}}{2m_{\mathcal N}} \right)^2 \frac{m^3_{\rho ^{(k)}}}{3M^2_{KK}} [2y_0(m_{\rho}^{(k)} r)] \\
  & + \sum _{k=1}^{P} \frac{1}{4\pi} (g_{a^{(k)}\mathcal{NN}})^2 \frac {m_{a^{(k)}}}{3} [-2  y_0(m_{a^{(k)}}r)],
\end{split}
\end{equation}
and
\begin{equation}\label{eq13}
\begin{split}
   V_T^S(r) = & \frac{1}{4\pi} \left( \frac{g_{\pi \mathcal{NN}}M_{KK}}{2m_{\mathcal N}} \right)^2 \frac{1}{M^2_{KK}r^3} \\
   & + \sum _{k=1}^{P} \frac{1}{4\pi} \left ( \frac{g_{\rho^{(k)}\mathcal{NN}}M_{KK}}{2m_{\mathcal N}} \right)^2 \frac{m^3_{\rho ^{(k)}}}{3M^2_{KK}} [-y_2(m_{\rho}^{(k)} r)] \\
   & + \sum _{k=1}^{P} \frac{1}{4\pi} (g_{a^{(k)}\mathcal{NN}})^2 \frac {m_{a^{(k)}}}{3} [ y_2 (m_{a^{(k)}}r)].
\end{split}
\end{equation}

\section{Holographic Deuteron}

Deuteron which contains one proton and one neutron is the only bound state of two-nucleon system. Deuteron only has the ground state with the isospin $T=0$, total spin $S=1$, spin parity $1^+$, and binding energy $E_B=-2.2246MeV$, and no excited states. Because the spin-parity of deuteron is $1^+$, the values of spin and isospin interactions can be determined using the superselection rules
\begin{equation}\label{eq23}
  S_{12}=2, \quad \vec{\sigma}_1\cdot\vec{\sigma}_2 = 1, \quad \vec{\tau}_1\cdot\vec{\tau}_2 = -3 .
\end{equation}

\begin{center}
\includegraphics[width=7.0cm]{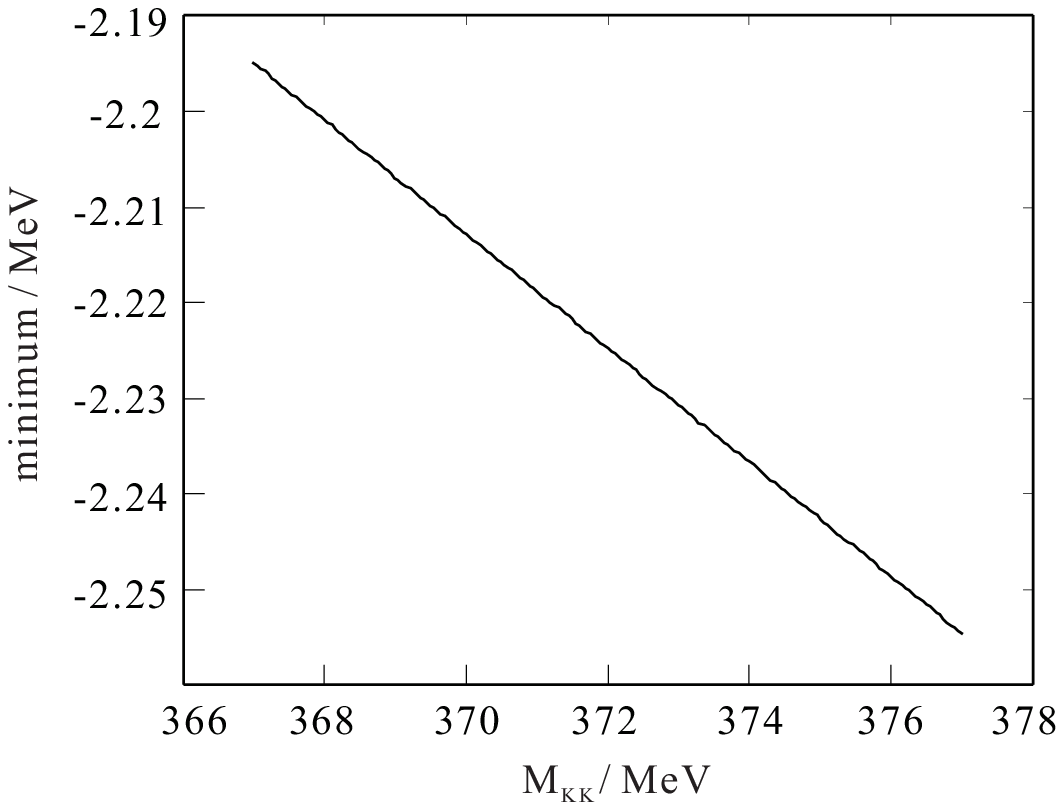}
\figcaption{\label{fig1}   The potential minimum of different $M_{KK}$. }
\end{center}

The $AdS_6$ model has just one free parameter which is the mass scale of the model, $M_{KK}$, and all the other parameters are derived from model calculations \cite{lab12}. The minimum of the deuteron ground potential is considered the deuteron binding energy. The minimum has been calculated as functions of the $M_{KK}$ and shown in Fig. \ref{fig1}. We can see that there is a basically linearity relationship between the minimum of the potential and the $M_{KK}$. The minimum is approximately equal to $¨C2.2246 MeV$ when the $M_{KK}$ is equal to $372 MeV$. The values of $N_c = 3$, $m_N = 920 MeV$, and $P=10$ are choosed in the calculations. Fig. \ref{fig2} shows the deuteron potential in terms of $rM_{KK}$ for the $M_{KK}=372MeV$. As it is indicated, the potential has repulsive behavior at short distances and becomes roughly zero at large $rM_{KK}$. It contains a shallows minimum in depth $\sim 2.225 MeV$ around $rM_{KK} = 7.62$.

\begin{center}
\includegraphics[width=6.0cm]{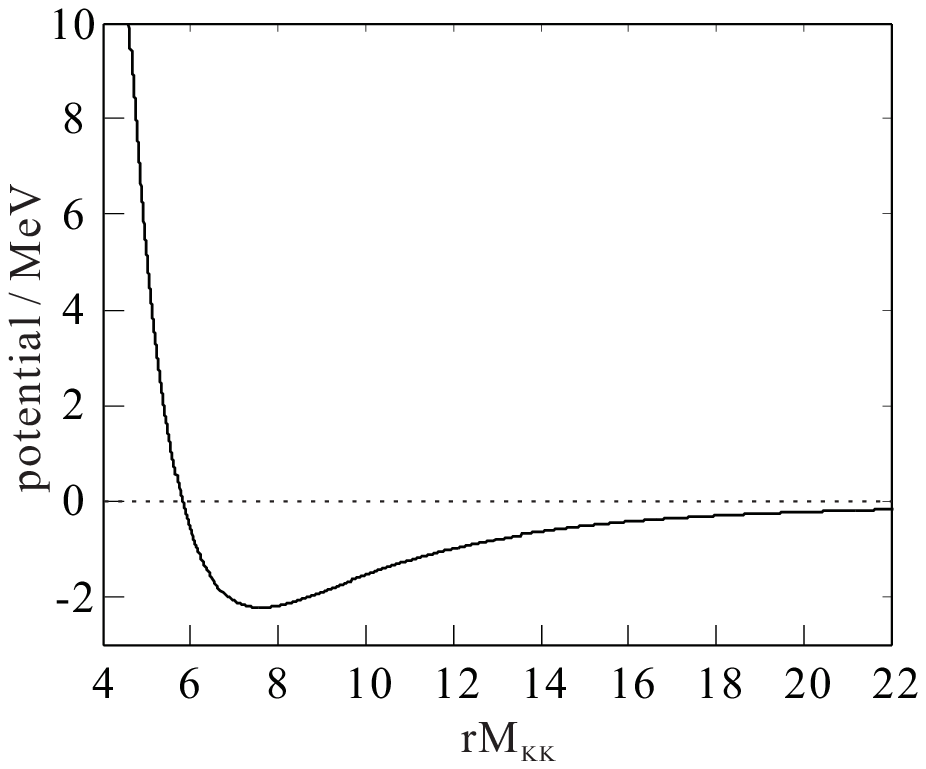}
\figcaption{\label{fig2}   The deuteron potential in terms of $rM_{KK}$ for the $M_{KK}=372MeV$. }
\end{center}

Fig. \ref{fig3} depicts the holographic potentials of the (a) pseudoscalar meson $\pi$, (b) isospin singlet vector mesons $\omega^{(k)}$, (c) isospin triplet vector mesons $\rho ^{(k)}$ and (d) the triplet axial-vector mesons $a^{(k)}$. We can see that the long-range part of the holographic deuteron potential is mostly due to the one pion exchange mechanism while the isospin singlet vector mesons $\omega^{(k)}$ produce the strong short-range repulsion. Exchanging the isospin triplet vector meson $\rho ^{(k)}$ can produce a small short-range repulsion and exchanging the triplet axial-vector mesons $a^{(k)}$ can produce the small attractive behavior for deuteron in the $AdS_6$ model.

\vspace{0.5cm}
\end{multicols}
\ruleup
\begin{center}
\includegraphics[width=11cm]{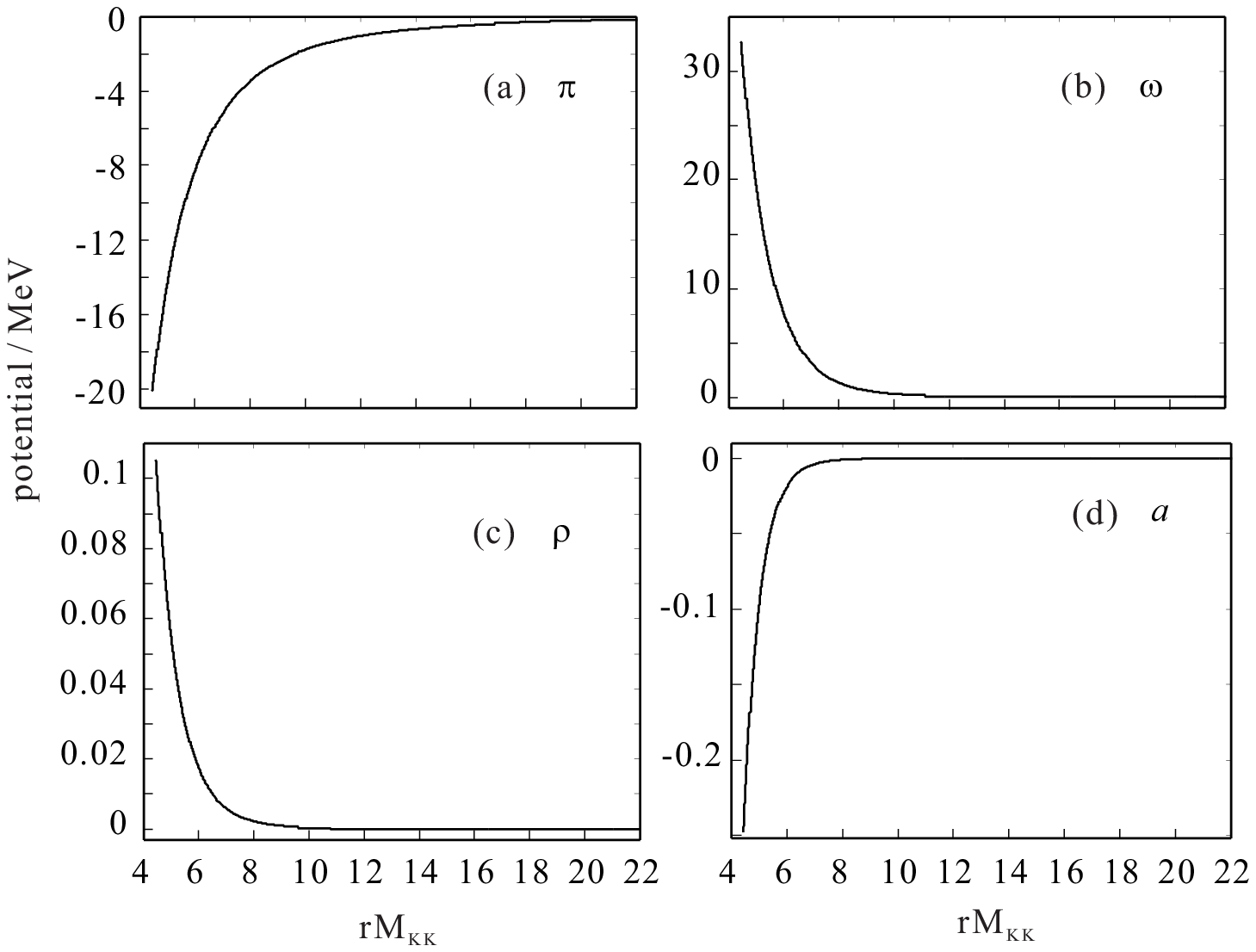}
\figcaption{\label{fig3} The holographic potentials of the (a) pseudoscalar meson $\pi$, (b) isospin singlet vector mesons $\omega^{(k)}$, (c) isospin triplet vector mesons $\rho ^{(k)}$ and (d) triplet axial-vector mesons $a^{(k)}$.}
\end{center}
\ruledown

\vspace {0.5cm}
\begin{multicols}{2}

Fig. \ref{fig4} depicts the potentials of different $k$ for the relevant part of summation in the deuteron holographic potential. It can be seen that the lowest state of various mesons play more major roles in the potential. The effect on the potential for $k+1$ nearly 2 order of magnitude lower than that for $k$ in the short-range part. So it is enough to choose $P=10$ in the calculation of the deuteron holographic potential.

\end{multicols}
\vspace{0.5cm}
\ruleup
\begin{center}
\includegraphics[width=11cm]{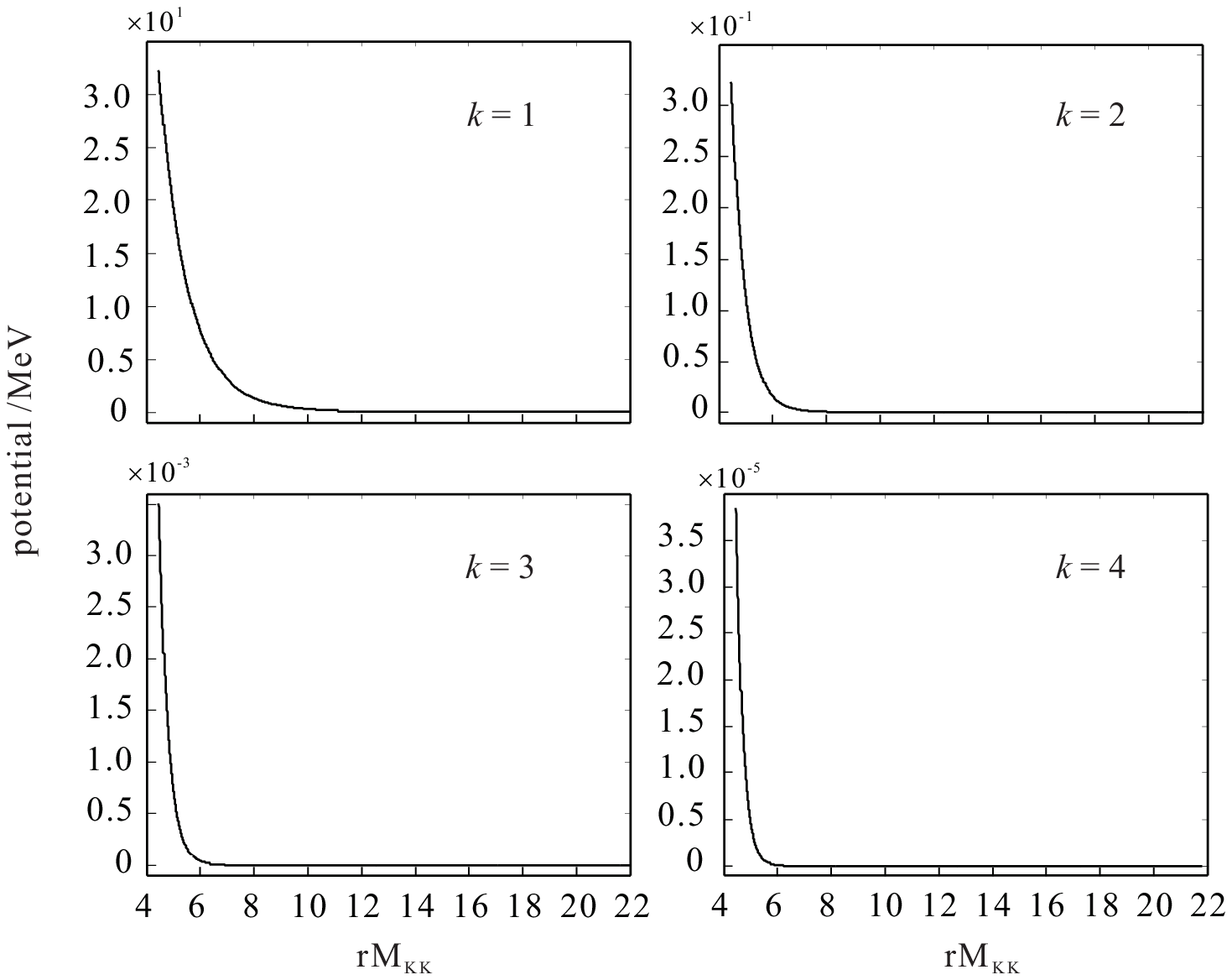}
\figcaption{\label{fig4} The potentials of different $k$ for the relevant part of summation in the deuteron holographic potential.}
\end{center}
\ruledown

\vspace{0.5cm}
\begin{multicols}{2}

\section{Conclusion}

Deuteron has been studied using a noncritical holographic QCD model constructed in the $AdS_6$ supergravity background on the basis of one-boson exchange picture. The deuteron potential has been calculated. The minimum of the deuteron potential is considered as the deuteron binding energy. Various mesons play the different roles in producing the NN interaction. The lowest state of various mesons play more major roles. This model has just one free parameter which is the mass scale of the model, $M_{KK}$, and all the other parameters are derived from model calculations [11]. So it has more clear physical image compared with some modern high-quality phenomenological NN interaction models.

\end{multicols}

\clearpage

%\end{CJK*}
\end{document}